\documentclass[pra,twocolumn,showpacs,preprintnumbers,superscriptaddress]{revtex4-1}

\usepackage{times}
\usepackage{subfig}
\usepackage{bm}
\usepackage{graphicx}
\usepackage{amsbsy}
\usepackage{amsmath}
\usepackage{amsfonts}
\usepackage{amsthm}
\usepackage{float}
\usepackage{color}
\DeclareMathOperator{\Tr}{Tr}
\DeclareMathOperator{\cov}{Cov}

\theoremstyle{plain}

\newtheorem{lemma}{Lemma}

\newtheorem{observ}{Observation}

\begin{document}

\voffset  0.5in

\title{Local, nonlocal quantumness and information theoretic measures}
\author{Pankaj Agrawal}
\thanks{agrawal@iopb.res.in}
\affiliation{Institute of Physics, Sainik School Post, 
Bhubaneswar-751005, Orissa, India.} 

\author{Sk Sazim}
\thanks{ssazim@iopb.res.in}
\affiliation{Institute of Physics, Sainik School Post, 
Bhubaneswar-751005, Orissa, India.} 
\author{Indranil Chakrabarty}
\affiliation{
Center for Security, Theory \& Algorithmic Research, International Institute for Information Technology, Hyderabad, India
}
\author{Arun K. Pati}
\affiliation{
 Harish-Chandra Research Institute, Chhatnag Road, Jhunsi, Allahabad 211 019, India
}

\date{\today}


\begin{abstract}
     \noindent It has been suggested that there may exist quantum correlations
      that go beyond entanglement. The existence of such correlations
      can be revealed by information theoretic quantities such as quantum discord, 
      but not by the conventional measures of entanglement. We argue that a state displays
      quantumness that can be of local and nonlocal origin. Information theoretic
      measures not only characterize the nonlocal 
      quantumness but also the local quantumness, such as the ``local superposition''. 
      This can be a reason why such measures are non-zero when there is 
      no entanglement.
      We consider a generalized version of the Werner state to demonstrate
      the interplay of local quantumness, nonlocal quantumness, and 
      classical mixedness of a state.

\end{abstract}

\pacs{}
\maketitle
\section{Introduction}
 One of the basic problems in quantum physics is to understand the nature of 
correlations present between different particles in a composite system. The existence 
of non-factorizable states play important role in the existence of 
many exotic features of quantum information theory.  These features, more specifically the 
advantages, include phenomena like 
quantum cryptography \cite{gisin}, quantum computation \cite{ani,mixcomp}, quantum imaging \cite{image},
quantum phase transition \cite{qphase}, quantum biology \cite{engel} and 
many more \cite{others}. 
Therefore, it is important to study and understand the nature of correlations present 
in various quantum systems. In the last decade, various measures of 
correlations \cite{entrev,OLI,REST,IND,raja,witn} have been introduced. 
It is believed that none of these measures can alone be sufficient to manifest all 
the  facets of quantum correlations. However, each of these measures unveils 
some aspects of quantum correlations.

 Entanglement is the key concept which alters the notion of reality in the microscopic 
 system. Not only that, it is also responsible for the metamorphosis of the 
 meaning of correlation as we move from classical systems to quantum systems. 
 Various studies have been conducted regarding the detection 
 as well as quantification of entanglement \cite{lew,ter,spe}. Multitudinous techniques and 
 measures have been introduced for the detection and quantification of entanglement. For pure states the 
 situation is quite comprehensible as entanglement tells all about the correlation present. 
 However, the situation is not so clear in the case of mixed states. In the case of mixed states 
 because of certain issues,  researchers begin to have a hunch that there 
 may be something beyond the entanglement that actually quantifies the amount of correlation 
 present in the system.  Recently, many measures have been proposed to quantify the amount of 
 correlation present in a mixed state. These measures have a unique feature that all of 
 them seem to predict the presence of quantum correlation beyond the domain of entanglement.
 However, the nature of these correlation, as one goes beyond the entanglement, is far from
 understood. 
 There have been many speculations, but still there is no universal way by which one can 
 determine the total correlation present in the system \cite{totalcorr}.

 There are two aspects of the quantum mechanical formalism that play important role in quantum 
information processing.  The one aspect may be 
referred to as nonlocal quantumness. This is due to nonlocal superposition. This nonlocal 
superposition leads to entanglement. We will call the other aspect as local quantumness. The local 
quantumness appears due to local superposition of the states. Every extant correlation 
measures which is non-zero for separable states will show such quantumness. For the 
sake of illustration and simplicity, we will focus here on the information theoretic 
measure -- quantum discord.

Of the late, quantum discord have been accentuated in many works \cite{OLI,REST,IND}. 
This is an important information theoretic measure based on the mutual information. 
It is the difference between the total correlation and the classical correlation present 
in the system. Here in this work, we demonstrate that such quantities probe not only 
nonlocal quantumness but also local quantumness. That is the prime reason why such measures 
are non-zero for mixed states even when there is no entanglement present in the system.

The organization of the work is as follows. In the section II, we discuss the notion of 
quantum covariance to characterize the correlations. In the section III, we give a 
brief introduction to the quantum discord. In the section IV, we discuss the phenomenon
of local and nonlocal quantumness. In the section V, we consider few states to exemplify 
the difference between the classical and separable states in the context of local and 
nonlocal quantumness. 
In the section VI, we introduce the parametric representation of local and nonlocal 
quantumness and show that the discord function depends on both of them. Finally, we conclude in 
the last section.

\section{Quantum Covariance}

The covariance for a bipartite state $\rho_{XY}$ is defined as
\begin{eqnarray}
\cov(\rho_{XY},{\cal O}_X,{\cal O}_Y) = \rm{Tr}_{XY}(\rho_{XY}{\cal O}_X {\cal O}_Y) - \nonumber \\
  \rm{Tr}_{X}(\rho_{X}{\cal O}_X) \rm{Tr}_{Y}(\rho_{Y}{\cal O}_Y),
\end{eqnarray}
where ${\cal O}_X$ and ${\cal O}_Y$ are observables acting on the part $X$ and $Y$ respectively. 
Unlike its classical counterpart, this covariance is not a measure of quantum entanglement 
(or quantum correlations). However, we can use it to detect quantum correlations. (See also
the discussion below about nonlocal quantumness.) Using the intuitive meaning of quantum
correlation, one can argue that a bipartite pure state has no quantum correlations, if the
covariance vanishes for any two arbitrary observables $X$ and $Y$.
Clearly, covariance vanishes for the product states $\rho_{XY} = \rho_{X} \otimes \rho_{Y}$. 
Here $\rho_{X}$ and $\rho_{Y}$ are reduced density matrices.   
For a mixed state, one can minimize the magnitude of quantum covariance over all possible 
decompositions. We can then define covariance for the system  with 
the density matrix $\rho_{XY} = \sum_{i} p_{i} \rho^{i}_{XY}$ as
\begin{equation}
 \varLambda(\rho_{XY})=\min \sum_{i} p_{i} |\cov(\rho^{i}_{XY},{\cal O}_X,{\cal O}_Y)|.
\end{equation}
To avoid the negative value of covariance, we have considered its magnitude.
In case the $\varLambda(\rho_{XY},{\cal O}_X,{\cal O}_Y)$ is non-zero, then the state
will have quantum correlations. 

\noindent{\lemma For all bipartite two-qubit separable states, $\varLambda(\rho_{XY})=0$.}

\noindent\textit{Proof:} Here we can use the fact that (a) all the separable states can be decomposed in terms of 
product states and (b) for product states $\varLambda=0$.

Hence the `Lemma 1' is important to identify bipartite correlated states.

\section{Quantum Discord}
The quantum discord was introduced by Olivier and Zurek (2002) \cite{OLI} as a measure of the 
 ``quantumness of correlations''. It is defined in terms of the mutual information. 
 Classically, the mutual information is a measure of common information in two 
 random variables. Therefore, it was natural to generalize it to the quantum 
 domain and express quantum correlation in terms of this object. However, the 
 definition of the mutual information in quantum domain is not straightforward. This is 
 because there are more than one classical expressions to define the mutual 
 information. These different expressions admit different generalizations. 
 Discord uses this difference to characterize the quantum correlations.
Classically, one can write the mutual information in two alternate ways,
$I(X:Y) =  H(X) - H(X|Y)$, and  
$J(X:Y)  =  H(X) + H(Y) - H(X,Y)$.
Here $H(X), H(X,Y)$ and $H(X|Y)$ are the entropy, joint entropy, and 
conditional entropy for the random variables $X$ and $Y$. The Joint 
entropy and conditional entropy are related by the chain rule, 
$ H(X|Y)=H(X,Y)-H(Y)$.

These expressions for the entropies can be generalized to the 
quantum domain by substituting random variables by density matrices 
and Shannon entropies by von Neumann entropies. For example,
$H(X) \to H(\rho_X)= - {\rm Tr}[\rho\log(\rho)].$
The generalization of the mutual information will also involve the generalization of 
the conditional entropy. We use the generalization as suggest in the Ref \cite{OLI}. 
Using this generalization to the quantum domain, we obtain
$I(X:Y)=H(X)-H(X|\{\pi^Y_i\})$,
where $H(X|\{\pi^Y_j\})=\sum_j p_jH(\rho_{X|\pi^Y_j})$ with $\rho_{X|\pi^Y_j}=\frac{\pi^Y_j \rho_{XY} \pi^Y_j}
{\Tr(\pi^Y_j \rho_{XY})}$ (where $p_j$ is the probability of obtaining the $j$th outcome).
Here,  $H(X|\{\pi^Y_j\})$ is the Von Neumann entropy of the qubit $X$, when the projective measurement is done on $Y$. The quantum discord is then defined as,
$D(X:Y)= J-I = H(Y)-H(X,Y)+ H(X|\{\pi^Y_j\}).$
This is to be minimized over the set of all one dimensional projectors $\{\pi^Y_i\}$. We shall call 
 $D(X:Y)$ as discord function and its minimum value as the quantum discord. It is evident that the discord 
 function is not symmetric in $X$ and $Y$. In the above definition, we are making a measurement
 on the system $Y$. Let us call it $Y$-discord. Similarly, we can define $X$-discord, when the measurement
 is made on the system $X$, $D(Y:X)= H(X)-H(X,Y)+ H(Y|\{\pi^X_j\})$.
Here $H(Y|\{\pi^X_j\}$ is defined in the same way as $H(X|\{\pi^Y_j\})$. 

For a bipartite state, $X$-discord and $Y$-discord may have different values. They will have 
 identical values when the state is symmetric in $X$ and $Y$. But, they are always non-negative. 
 When one of the discord is zero, then the state would be separable. However it still 
 may not be completely classical state and may exhibit quantum behaviour. For the state 
 to be completely classical, both discords must vanish. As we shall see below, there 
 exist states for which only one of these discords is zero. Therefore, for the complete 
 characterization of the quantumness, one should know both discords.
For our convenience, we define 
a vector quantity, $\vec{\delta}$ which contains both discord as,
\begin{equation}
 \vec{\delta}(\rho_{XY})=\{\delta(X:Y),\delta(Y:X)\}.
\end{equation}
where $\delta(X:Y)$ and $\delta(Y:X)$ are the $X$-discord and $Y$-discord respectively after minimization over measurement parameters. 

\noindent{\observ A two-qubit state is either classically correlated or is a product 
state iff $\vec{\delta}=\vec{0}$.}

In the literature, there exit witness operators for discord (cf. \cite{witn}) but we will 
not discuss them here. Observation \textbf{1} is enough for our analysis and it 
also gives us information about the structure of the states.

\section{Quantumness -- local and nonlocal}

A state of a bipartite quantum system may exhibit nonclassical behaviour
  due to either the local superposition (``local quantumness'') or due to
  the nonlocal superposition, i.e. entanglement, (``nonlocal quantumness'').
  Usually, one is more concerned about the entanglement and its
  characterization and quantification -- in part due to its mysterious
  nature and to use it as a resource. However, local quantumness
  can also be important if we can exploit the superposition as a
  resource in general. It is the superposition, local or nonlocal, that gives
  advantage in many quantum information processing protocols.

  In the case of quantum discord, therefore we have $D(X:Y) = D(\varphi_{L}(X:Y), \varphi_{NL}(X:Y))$
  where $\varphi_{L}(X:Y)$ characterizes the local quantumness, and
  $\varphi_{NL}(X:Y)$ characterizes the nonlocal quantumness of the state.
  We don't know yet if  $D(X:Y) = D(\varphi_{L}(X:Y) + D(\varphi_{NL}(X:Y)$.
  Their properties are -- 1) both $\varphi_{L}(X:Y)$ and $\varphi_{NL}(X:Y)$ are
  invariant under local unitary operation; 2) $\varphi_{L}(X:Y)$ may increase
  under local operations, but  $\varphi_{NL}(X:Y)$ would not; 3) under global 
  operations, $D(X:Y)$ may increase or decrease; 4) if the state is separable
  then  $D(X:Y) = D(\varphi_{L}(X:Y))$ and $D(X:Y) = D(\varphi_{L}(X:Y), \varphi_{NL}(X:Y))$ 
  for an entangled state. 
 We now discuss these features of a quantum state in a bit more detail.

\subsection{Local Quantumness}

A bipartite separable quantum state may not have entanglement,
   but it is a quantum state and can exhibit quantum features. This
   quantum feature may be called as the local quantumness.
   What we mean by local quantumness can be seen by
   following three examples. Consider the density operators
\begin{eqnarray}
\rho_{a} & = & |++\rangle\langle ++|,  \nonumber \\
\rho_{b} & = & p\, |++\rangle\langle ++|\;\; +\;\; (1-p) \, |--\rangle\langle --|, \nonumber \\
\rho_{c} & = & q\, |++\rangle\langle ++|\;\; +\;\;(1- q)\, |00\rangle\langle 00|, 
\end{eqnarray}
where $ |\pm\rangle = {1 \over \sqrt{2}} ( |0\rangle \pm  |1 \rangle)$. 
The state $\rho_a$ is a pure quantum state with no entanglement. Now one can argue that 
this state shows local quantumness. However, this local quantumness can be
 masked in the case of a pure product state.  If we make a local 
 measurement on the particle 
 `$A$' in Hadamard basis, we will get the particle in the state $|+\rangle$ 
 with unit probability and the state would not change after the measurement.
 So, the local quantumness may not apparent.
 However, if we make a measurement in the computational basis 
 $\{ |0\rangle,  |1\rangle  \}$, then
 the particle `$A$' can be found in any of the computational basis state with
 equal probability and the state would change after the measurement.
 This can be easily seen if we think of the state as a local superposition 
 of the computational basis states.  
 The state $\rho_b$ is what is known as classical mixed state. Its
          behavior will be similar to $\rho_a$ with respect to the measurements.
          We can mask its local quantumness. The state $\rho_c$ is also a separable 
          state. However, in this state we
          cannot mask the local quantumness, irrespective of the measurement
          basis. This is because one particle state is not orthogonal and the 
          state in one of mixture component can be written in terms of the
          superposition of the state in the other component and the rest of the
          measurement basis. Therefore, irrespective of the measurement basis,
          local quantumness  (local superposition) cannot be hidden. 
 So, we see that a separable state which is not completely classical, will
 have local superposition which can be exploited. This is what has been showing
 up as a resource in the case of, eg, the model deterministic quantum computational
 with one quantum bit (DQC1) \cite{ani}.

\noindent{\lemma A two qubit state $\rho_{XY}$ has only local quantumness iff $\varLambda(\rho_{XY})=0$ and $\vec{\delta}(\rho_{XY})\neq \vec{0}$.}

\noindent{\it Proof:} The proof follows from the observations (a) for all two-qubit separable 
states $\varLambda(\rho_{AB})=0$ and, (b) only for product states, or, classical states 
$\vec{\delta}(\rho_{XY}) = \vec{0}$. 
   
\subsubsection{Local noise can enhance Discord}

    Since discord probes also local quantumness, therefore it can 
    even increase by local operations. However, the local operation
    should be such that it changes the relative local quantumness 
    of the mixture components. Quantum noise can be a good candidate
    for such a local operation. However standard local noise such
    as bit flip and phase flip noise cannot change discord, because
    no relative local superposition is introduced. In Ref \cite{skb},
    a set of Krass operators are given which can convert a classical
    mixed state, like $\rho_1$, given below in Eq. (\ref{example_dis}), to a classical-quantum mixed state,
    like $\rho_2$ or $\rho_3$, given below. This local noise can convert
    one separable state to another separable state, but not to an
    entangled state. This noise is only changing the local quantumness
    properties of a bipartite state.

\subsection{Nonlocal Quantumness}

  In this paper, we shall mean the existence of quantum correlations
  in a state as equivalent to the state showing ``nonlocal quantumness''.
  It will also be synonymous with the existence of entanglement.
  If there is a system
  made of two subsystems, and there are quantum correlations, then 
  the properties of the one subsystem, say $A$,
  would depend on the properties of the other subsystem, say $B$.
  The states of the subsystems are interdependent.
  This is the intuitive meaning of correlations. 
   One can give a criteria for a pure bipartite state 
   to possess quantum correlations. This criteria can then be
   generalized to a mixed state. This has been discussed in the section II.

\noindent{\observ A two qubit state $\rho_{XY}$ has nonlocal quantumness iff $\varLambda(\rho_{XY})\neq 0$.}

 This just follows from the Lemma 1.

We have discussed above the quantumness of a state goes beyond
   entanglement. We suggest that discord characterizes the quantumness of a state. 
   This quantumness has both local and nonlocal components. A separable state can have 
   local quantumness, but no nonlocal quantumness. An individual system may also
   show quantum, i.e., non-classical behaviour. So quantum behaviour of any system 
   encompasses quantumness due to correlation and quantumness of an individual 
   system in the absence of correlation. Essence of the local quantumness is due 
   to the superposition property of the state of a subsystem of a composite system. 
   We can visualize the classification of bipartite states as in Fig 1.

\begin{center}
\begin{figure}[h]
\[
\begin{array}{cc}
\includegraphics[height=4.5cm,width=6.5cm]{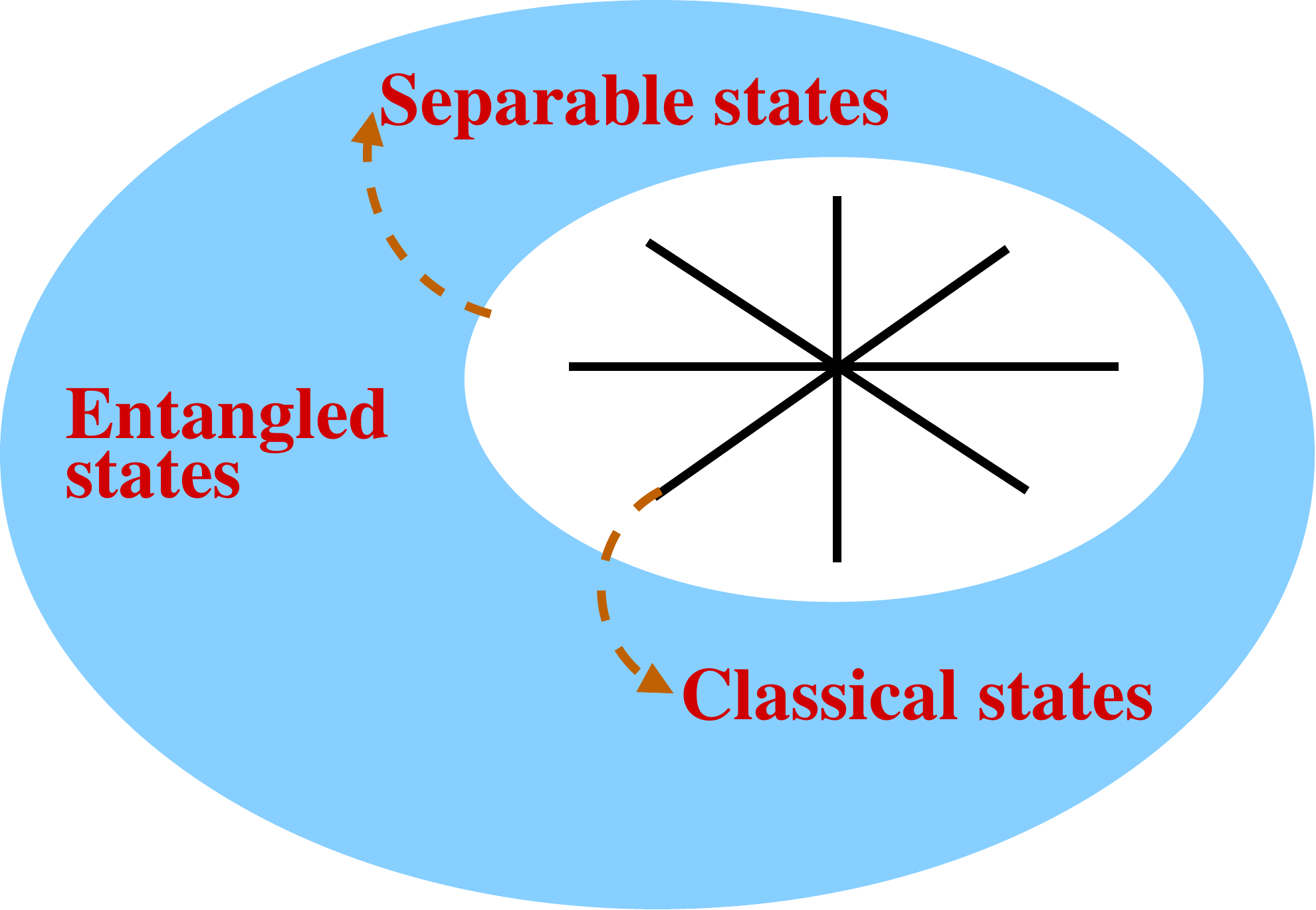}
\end{array}
\]
\caption{\noindent({\it Color online}) The large ellipse represents all two-qubit states (cf. \cite{sta} ). The small ellipse represents all separable states (i.e., $\varLambda=0$). The lines represent set of product states (end points of the lines) and classical states (in different basis). The point where the lines meet is the maximally mixed state. The outer annular space contains all entangled or, nonlocal states (i.e., $\varLambda\neq 0$ \& $\vec{\delta}\neq \vec{0}$), inner ellipse (except the lines) contains all separable states with local quantumness (i.e., $\varLambda= 0$ \& $\vec{\delta}\neq \vec{0}$ ) and line depicts all product states and classical states ($\varLambda= 0$ \& $\vec{\delta}=\vec{0}$).}
\end{figure}
\end{center}

\section{Simple Examples}

\subsection{Separable and Classical States}

In order to exemplify our argument, we consider separable mixtures 
and examine the discord function for them. In the mixed state domain, a state is 
said to be separable if it can be expressed as convex combination of 
product states.  So, in principle a product state is a separable state 
while the converse is not always true. Therefore, separable states do not possess
entanglement. However, as is known, not all separable states have zero discord.
As a paradigm, we start with following mixed states,
\begin{eqnarray}
\rho_{1} & = & p\, |00\rangle\langle 00| \;\; +\;\; (1-p) \,|11\rangle\langle 11|, \nonumber \\
\rho_{2} & = & p\, |++\rangle\langle ++| \;\; +\;\; (1-p) \,|0-\rangle\langle 0-|, \nonumber \\
\rho_{3} & = & p\, |++\rangle\langle ++| \;\; +\;\; (1-p) \,|-0\rangle\langle -0|, \nonumber \\
\rho_{4} & = & p\, |++\rangle\langle ++|\;\; +\;\; (1-p) \, |00\rangle\langle 00|, 
\label{example_dis}
\end{eqnarray}
where $ |\pm\rangle = {1 \over \sqrt{2}} ( |0\rangle \pm  |1 \rangle)$ are the Hadamard states.
 These density matrices represent four different categories of separable states. Neither of these states have entanglement. 
However, these states differ in important ways. $\rho_1$ belongs to the category of completely 
classical states. $\rho_2$ and $\rho_3$ are not completely classical, because, in the mixture,  
the states of only one of the particles are orthogonal. In the case of $\rho_1$, both 
$X$-discord and $Y$-discord are zero. For $\rho_2$, $X$-discord is zero, while for 
$\rho_3$, $Y$-discord is zero. For $\rho_4$, both discords are non-zero.
If we make a measurement in computational basis, then the discord function 
is nonzero for $\rho_1$ and $\rho_2$. But we have to minimize the discord 
function to obtain the discord, the discord is zero for $\rho_1$, 
but not for $\rho_2$. For $\rho_1$, the discord function is zero in the 
Hadamard basis. This is the basis formed out of the states, of which 
the $\rho_1$ is a mixture. In this basis conditional entropy is zero, 
while entropies of the individual and composite system cancel. 
These facts are illustrated in Figs 1 and 2, where discord functions
are plotted as a function of the angle $\theta$ that characterizes the
measurement basis \cite{IND}. In these plots, $DX = D(Y:X)$ and $DY = D(X:Y)$.
\begin{widetext}
\begin{center}
\begin{figure}[h]
\[
\begin{array}{cc}
\includegraphics[height=4.0cm,width=5.5cm]{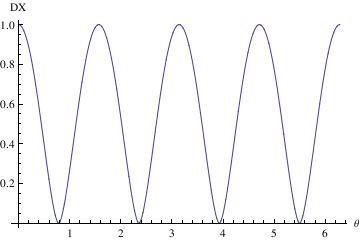}&
\includegraphics[height=4.0cm,width=5.5cm]{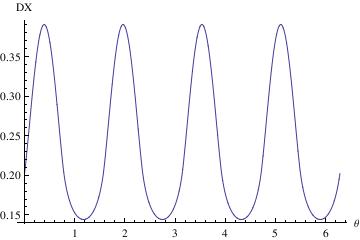}\\
(i)&(ii)
\end{array}
\]
\caption{ Dependence of $X$-Discord function on measurement basis with classical mixing
parameter $p$ = 0.5 for (i) $\rho_1$ and (ii) $\rho_4$.}
\end{figure}
\end{center}
\begin{center}
\begin{figure}[h]
\[
\begin{array}{cc}
\includegraphics[height=4.0cm,width=5.5cm]{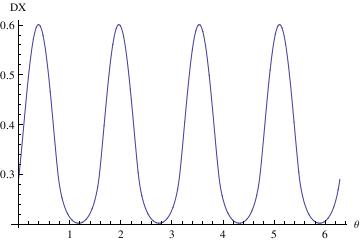}&
\includegraphics[height=4.0cm,width=5.5cm]{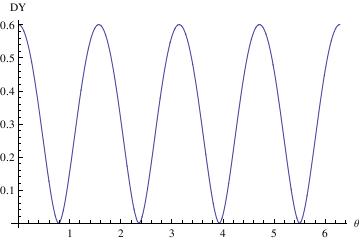}\\
(i)&(ii)
\end{array}
\]
\caption{ Dependence of (i) $X$-Discord  and (ii) $Y$-Discord
functions on  measurement basis with classical mixing parameter $p$ = 0.5 for $\rho_2$. For $\rho_3$
the $X$-Discord and $Y$-Discord  are interchanged.}
\end{figure}
\end{center}
\end{widetext}
This is in accordance with the fact that while the density operator $\rho_1$ 
represents a classical mixture, i.e., a mixture of orthogonal states, 
the density mixture $\rho_4$ represents a mixture of non-orthogonal 
states. In the case of $\rho_4$, unlike $\rho_1$, states in one of the component, 
$|+\rangle$ is a linear superposition of the computational basis states $\{|0\rangle,
|1\rangle \}$. This is the case of local superposition. Therefore, the discord 
is non-zero for $\rho_4$ because it also probes local quantumness (apart from  
nonlocal quantumness due to entanglement). One can say that a mixture of 
non-orthogonal  separable state has local quantumness, i.e., local superposition, 
which cannot be washed away by writing down another decomposition of the density matrix.

\subsection{Werner State}
In this subsection, we show the importance of local quantumness for non vanishing value of
 the quantum discord with the aid of the Werner state. This state is given by
\begin{equation}
        \rho_{\rm w} = (1-p)\,\frac{\mathbb{I}}{4}\;\;+\;\;p\,| \Phi^+\rangle\langle \Phi^+|,
\end{equation}    
    where $|\Phi^+\rangle\ = {1 \over \sqrt{2}}  (|00\rangle + |11\rangle) $ is a
     Bell state, $\mathbb{I}$ is the identity operator and $p$ is the classical mixing
     parameter. Naively, one may think that this state is not separable and
    has quantum correlations for all values of classical mixing parameter
    $p$. However, it is known that this state is not entangled when
    $p < {1 \over 3}$ (using Peres-Horodecki criterion \cite{ph}, e. g.).
 It is also known that this state (pseudo pure state) is useful
    for information processing.
 If we look at the plot of the discord and the concurrence in the Fig 3, we see
    that concurrence is zero, when the state is not entangled, but
   the discord is non-zero.
\begin{widetext}
\begin{center}
\begin{figure}[h]
\[
\begin{array}{cc}
\includegraphics[height=4.0cm,width=5.5cm]{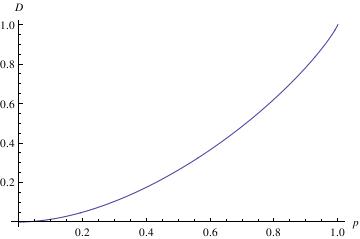}&
\includegraphics[height=4.0cm,width=5.5cm]{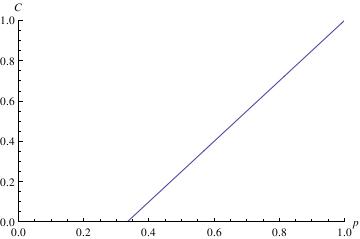}\\
(i)&(ii)
\end{array}
\]
\caption{ (i) Quantum discord (D)  and (ii) Concurrence (C) for the Werner State as a function of
    the classical mixing parameter $p$.}
\end{figure}
\end{center}
\end{widetext}
   At this point we ask this question:  Does it necessarily mean that Werner state has quantum 
   correlations that, in some sense, go beyond entanglement ? We claim that the answer to 
   this question is no. Our argument is that one can always rewrite Werner state in such a 
   way that this state is a valid mixture of non-orthogonal states whenever  $p < {1 \over 3}$ \cite{raja}. 
   Therefore the discord is nothing but just revealing the local quantumness. Rewriting the Werner state 
   in that form, we have
\begin{eqnarray}
 &&\rho_{\rm w}= (1 - 3 p)\,\frac{\mathbb{I}}{4}\;\;  
  +\;\;\frac{p}{2}\, (|++\rangle\langle ++| \;\;+ \;\;|--\rangle\langle --| \;\; {}\nonumber \\&&
|00\rangle\langle 00| \;\;+\;\; |11\rangle\langle 11|  \;\;  
+\;\; |\tilde{+}\tilde{-}\rangle\langle \tilde{+}\tilde{-} |\;\;
 + \;\; \ |\tilde{-}\tilde{+}|\rangle\langle |\tilde{-}\tilde{+}| ), \nonumber\\
\end{eqnarray}
 where $ | \tilde{\pm} \rangle\ = {1 \over \sqrt{2}}  (|0\rangle \pm i |1\rangle) $. This is a valid density operator when $ p \leq \frac{1}{3}$. This is 
 precisely the region of $p$, where Werner  state is not entangled.
 Since $ \langle +| 0 \rangle \neq 0$ and  $ \langle +| \tilde{+} \rangle \neq 0$, this 
 state is a mixture of separable non-orthogonal states; so it is expected
 to have non-zero discord due to local quantumness.

\section{Generalized Werner State: A comparative analysis of Entanglement and Discord}

In this section we generalize the Werner state to investigate the interdependence of 
local quantumness, nonlocal quantumness and classical mixedness by parametrization of 
each of these quantities. The major thrust of our claim lies in this part where 
we are able to see that the measures of entanglement like concurrence are independent 
of local quantumness, where as discord is a function of 
all these quantities. The generalized Werner state is defined as
\begin{equation}
\rho_{\rm GW} = (1-p)\,\frac{\mathbb{I}}{4}\;\;+\;\;p\,| \Phi_{nk}^+\rangle\langle \Phi_{nk}^+|,
\end{equation}
where, $| \Phi_{nk}^+\rangle  =  N_{nk}\,  (|+\rangle_{n} |+\rangle_{n} \; +\; k |-\rangle_{n} |-\rangle_{n} )$ ,
$|+\rangle_{n} = N \,(|0 \rangle + n  |1 \rangle) \;\;\;\;\; |-\rangle_{n} = N\, ( - n^{*} |0 \rangle +   |1 \rangle)$. Here $N_{nk}$ and $N$ are normalization constants. 
We can think of $n$ as a local superposition parameter; $k$ as a nonlocal superposition parameter 
and $p$ as the classical mixing parameter. We note that this state becomes a separable state as $k \to 0$. Furthermore,
there is no local superposition as $n \to 0$.
To study the behavior of the state with respect to these parameters,
we compute concurrence and discord for this state.
To see how the discord and concurrence change as we vary $p, n$ and $k$, in the following figures, we have plotted these functions.
In the Fig 4, we have plotted concurrence for two different values of $p$ as a function of the parameters
$n$ and $k$. We observe that concurrence is independent of the local superposition parameter $n$.
It is important because discord depends on $n$. It is expected that measures of entanglement are independent of 
local superposition parameter ($n$), while the measures of correlations which claim to go beyond entanglement will depend on it. 
Coming back to these, we see that concurrence vanishes when mixing is small and the state is not entangled. 
It is also noteworthy to see that larger the value of  $p$, larger is the concurrence.  Concurrence also vanishes 
when nonlocal superposition parameter is small. 
In other words, if one is small then other has to be large for the state to be entangled. 
In fact, we find that this generalized Werner state is entangled, i.e., the concurrence is non-zero when 
\begin{equation}
 p > {(1+k^2) \over (1 + k^2 + 4k)}.
\end{equation}
This requirement is independent of $n$. And it reduces to familiar  condition $ p > {1 \over 3}$ 
for the Werner state $(n = 0, k = 1)$ in order that it is entangled.

Let us now see how discord varies with respect to changes in $p, n$ and $k$. 
Similar to the concurrence, the discord is plotted in Figs 5 and 6.
With the increase of the value of mixing parameter, the value of 
discord increases. Even for very small values of mixing, when there is expected to be no entanglement, the discord is
non-zero. When there is no local superposition, i.e., $n=0$, the discord value increases as mixing becomes stronger. The value also 
increases, as the value of nonlocal parameter $k$ increases, i.e., entangled component of the mixture becomes more 
entangled, as expected. When there is no nonlocal superposition, i. e. $k=0$, and the mixture is separable, the discord is non-zero. Its 
value increases, as the mixing parameter increases, or local quantumness becomes stronger. Here important point is that the concurrence is independent 
of the local superposition parameter $n$, while the discord increases with an increase in the value of $n$.
\begin{widetext}
\begin{center}
\begin{figure}[h]
\[
\begin{array}{cc}
\includegraphics[height=4.0cm,width=5.5cm]{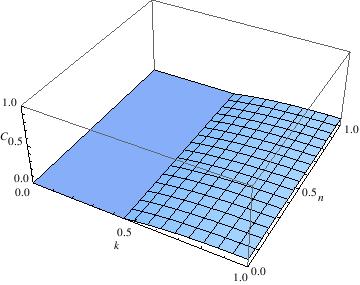}&
\includegraphics[height=4.0cm,width=5.5cm]{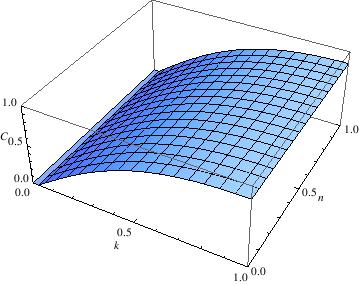}\\
(i)&(ii)
\end{array}
\]
\caption{Variation of concurrence (C) for the Generalized Werner State with local
superposition parameter $n$ and nonlocal superposition parameter $k$ for the
two values of classical mixing parameter (i) for $p$ = 0.4  and (ii) $p$ = 0.9.}
\end{figure}
\end{center}
\begin{center}
\begin{figure}[h]
\[
\begin{array}{cc}
\includegraphics[height=4.5cm,width=6.5cm]{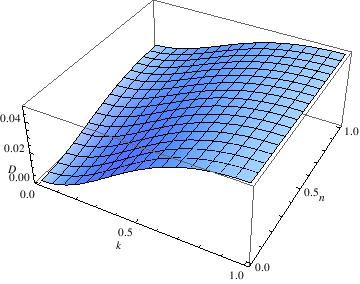}&
\includegraphics[height=4.5cm,width=6.5cm]{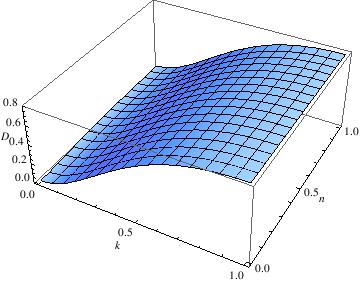}\\
\end{array}
\]
\caption{Variation of quantum discord (D) for the Generalized Werner State with local
superposition parameter $n$ and nonlocal superposition parameter $k$ for the
two values of classical mixing parameter (i) for $p$ = 0.2  and (ii) $p$ = 0.9.}
\end{figure}
\end{center}
\begin{center}
\begin{figure}[h]
\[
\begin{array}{cc}
\includegraphics[height=4.0cm,width=5.5cm]{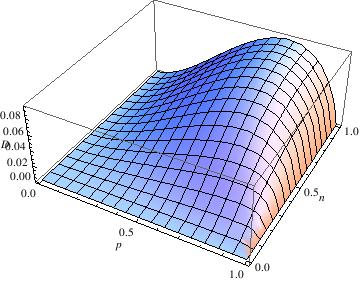}&
\includegraphics[height=4.0cm,width=5.5cm]{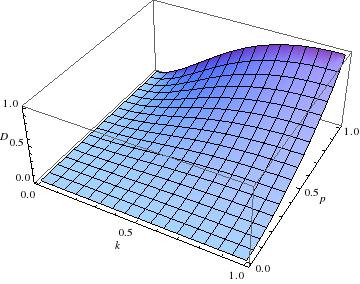}\\
(i)&(ii)
\end{array}
\]
\caption{Quantum discord (D) for the Generalized Werner State (i) for $k$ = 0 as 
a function of $p$ and $n$ and (ii) for $n$ = 0 as a function of $p$ and $k$.}
\end{figure}
\end{center}
\end{widetext}

       \section{Conclusion}

   We have proposed that quantum discord (and other similar measures)
   as a measure of quantum correlations for a bipartite system contains both 
   the local and the nonlocal quantumness. A quantum states with nonzero value 
   of discord does not mean existence of quantum correlations beyond entanglement. 
   In the absence of entanglement, there can be local quantumness that can 
   make the discord nonzero. We have illustrated our proposal using
   a generalized Werner state to demonstrate the interplay of local 
   quantumness, nonlocal quantumness, and classical
  mixedness by computing concurrence and quantum discord. To characterize the 
  quantumness of a state, one also needs to compute both X-discord
  and Y-discord. Both discords have to be zero to mask the local
  quantumness. 
 
We hope the present findings will help in understanding
  the nature of quantumness that goes beyond entanglement.

\textbf{Acknowledgement:} P. Agrawal thank the organizers of the event \textit{Quantum Discord Workshop, 2012}, CQT, Singapore for providing the platform to present this work and having fruitful discussions with the experts in the fields.

\end{document}